\documentclass[aps,graphicx,twocolumn,epsfig]{revtex4}

\usepackage{epsfig}

\begin{document}

\author{C. Menotti$^{1,2}$}
\author{C. Trefzger$^{1}$}
\author{M. Lewenstein$^{3}$}
\affiliation{$^1$ ICFO - Institut de Ciencies Fotoniques,
Mediterranean Technology Park, 08860 Castelldefels (Barcelona), Spain \\
$^2$ CRS BEC-INFM and Dipartimento di Fisica, Universit\`a di Trento,
I-38050 Povo, Italy\\
$^3$ ICREA and  ICFO - Institut de Ciencies Fotoniques,
Mediterranean Technology Park, 08860 Castelldefels (Barcelona), Spain}

\title{Metastable states of a gas of dipolar bosons in a 2D optical
lattice}

\begin{abstract}
We investigate the physics of dipolar bosons in a two dimensional
optical lattice. It is known that due to the long-range character of
dipole-dipole interaction, the ground state phase diagram of a gas of
dipolar bosons in an optical lattice presents novel quantum phases,
like checkerboard and supersolid phases.  In this paper, we consider
the properties of the system beyond its ground state, finding that it
is characterised by a multitude of almost degenerate metastable
states, often competing with the ground state. This makes dipolar
bosons in a lattice similar to a disordered system and opens
possibilities of using them as quantum memories.
\end{abstract}

\maketitle

The effect of long range interaction on the quantum phases of
ultracold gases in optical lattices has been recently investigated in
the literature \cite{santos,sinha}.  Theoretical studies have pointed
out that novel quantum phases, like supersolid and checkerboard
phases, arise as soon as the interaction potential involves at least
one nearest neighbour.  This issue has recently become of primary
importance, because, since the achievement of Bose-Einstein
condensation of dipolar Chromium atoms \cite{pfau} and with the
progress in cooling and trapping of dipolar molecules \cite{EPJD},
these systems are becoming to be at experimental reach.

As mentioned above, the phase diagram for non-zero range interactions
presents two main kinds of phases: (i) {\it superfluid supersolid
phases}, where
large atom-number fluctuations are found at each lattice site and
an order parameter different from zero characterises the system
\cite{batrouni,sengupta}; (ii) {\it insulating checkerboard phases},
where number fluctuations are absent and a well defined number of
atoms is found in each lattice site. These two phases differ from the
usual superfluid and Mott insulating phases, because they present
modulated patterns in the density and in the order parameter (when
different from zero). Usually these patterns consist in regular
distributions of atoms in the lattice sites, and are characterised by
a filling factor (average number of atoms per site) which is in
general not integer, even in the insulating phases.

The clear observation of such phases is a very important
experimental challenge. Although the observation of a possible
supersolid phase in $^4$He has been reported \cite{kimchan},
another proof of the existence of such quantum phases is desired.
We think that it could be eventually obtained with samples of
ultracold atomic gases in optical lattices in the presence of long
range interaction \cite{scarola}.

In this work, we focus our attention on the insulating states which
are found in the low-tunneling region of the phase diagram.  We point
out the existence of metastable states in the system. Contrary to the
standard on-site Bose-Hubbard model, we find that beyond the ground
state, there exist a huge amount of configurations (``classical''
distribution of atoms in the lattice sites), which have higher energy
but result stable against tunneling.  For small lattices ($4 \times
4$) and periodic boundary conditions, we analyse all possible existing
configurations and we find that for given chemical potential and
tunneling parameter there can be as much as hundreds of metastable
configurations.  We then generalize our results to large lattice sizes
(typically up to $20 \times 20$), comparable to the ones found in
realistic experimental situations.

Characterising systematically the metastable states in terms of
their stability against perturbations and their capability of
being approached in the time evolution of the system in the
presence of dissipation is important in view of the possible
application of those systems as quantum memories.
To our knowledge, the existence of metastable states has not been
discussed in the literature on extended Bose-Hubbard models so far
\cite{roscilde}.
Our results are based on a mean-field description of the system, but
we believe that the existence of the metastable states should be
confirmed by treatments beyond mean-field which are sensitive to local
minima of the energy.

We consider a two-dimensional (2D) gas of dipolar bosons in the
presence of a 2D optical lattice, and an extra confinement in the
perpendicular direction \cite{footnote0}.
We assume a single component gas of bosons (i.e. spin, or pseudo-spin,
polarised) \cite{sengupta2}.  Our system is well described by the
extended Bose-Hubbard Hamiltonian

\begin{eqnarray}
H &=& -\frac{J}{2} \sum_{\langle i j\rangle}\left(
 a^{\dag}_i a_j + a_i a^{\dag}_j \right) -  \sum_i \mu n_i +\\
&+&\sum_i \frac{U}{2}  \; n_i(n_i-1)
+  \sum_{\vec \ell}
\sum_{\langle \langle i
 j \rangle \rangle_{\vec \ell}} \frac{U_{\vec \ell}}{2} \; n_i n_j ,
\nonumber
\label{GBH}
\end{eqnarray}
where $J$ is the tunneling parameter, $U$ the on-site interaction,
$U_{\vec \ell}$ the components of the dipole-dipole interaction at
different relative distances, and $\mu$ the chemical potential which
fixes the average atomic density. The notation $\langle i j \rangle$
represents nearest neighbours and $\langle \langle i j \rangle
\rangle_{\vec \ell}$ represents neighbours at distance $\vec \ell$.

The on-site interaction is given by two contributions: one is arising
from the $s$-wave scattering $U_s= 4 \pi \hbar^2 a/m \int n^2(r) d^3
r$, and the second one is due to the on-site dipole-dipole interaction
$U_{dd}= 1/(2\pi) \int {\tilde V}(q) {\tilde n}^2(q) d^3q$, being
${\tilde V}(q)$ and ${\tilde n}(q)$ the Fourier transform of the
dipole potential and density, respectively \cite{goral}.
Due to the localisation of the wavefunctions at the bottom of the
optical lattice wells, the long range part of the dipole-dipole
interaction $U_{\vec \ell}$ is in a very good approximation given by
the dipole-dipole interaction potential at distance ${\vec \ell}$,
$U_{\vec \ell}= D^2 [1-3\cos^2(\alpha_{\vec \ell})]/\ell^3$,
multiplied by the densities $n_i$ and $n_j$ in the two sites. The
quantity $D$ is the dipole moment and $\alpha_{\vec \ell}$ is the
angle between the orientation of the dipoles and ${\vec \ell}$.

The ratio between the total on-site interaction $U=U_s+U_{dd}$ and the
nearest neighbour dipolar interation $U_{NN}$ determines much of the
physics of the system. It can be varied by tuning the on-site
dipole-dipole interaction $U_{dd}$ from negative to positive by
changing the vertical confinement, or by changing the $s$-wave
scattering length via a Feshbach resonance, as recently demonstrated
with Chromium atoms \cite{pfau07}. Alternatively one can consider to
use heteronuclear molecules, or Rydberg atoms, which possess much
larger dipole moments and will be hopefully soon available
experimentally in optical lattices.

The dipolar interaction potential decays as the inverse cubic power of
the relative distance. In most theoretical approaches the range is
cut-off at certain neighbours.  The precise choice of cut-off range
and lattice size, determines the fractional character of the allowed
ground state filling factors (e.g. for 1NN (or 2NN) only multiples of
1/2 (or 1/4) fillings are found).
In the present work, we consider a range of interaction up to the 4th
nearest neighbour and focus on the case of dipoles pointing
perpendicular to the plane of the lattice, where dipole-dipole
interaction between atoms in the plane of the lattice becomes
isotropic, and in particular always repulsive. We consider the case of
dipole-dipole interaction relatively weak ($U/U_{NN}=20$) and strong
($U/U_{NN}=2$) with respect to the on-site interaction
\cite{footnote}.

\begin{center}
\begin{figure}[t]
\includegraphics[width=0.95\linewidth]{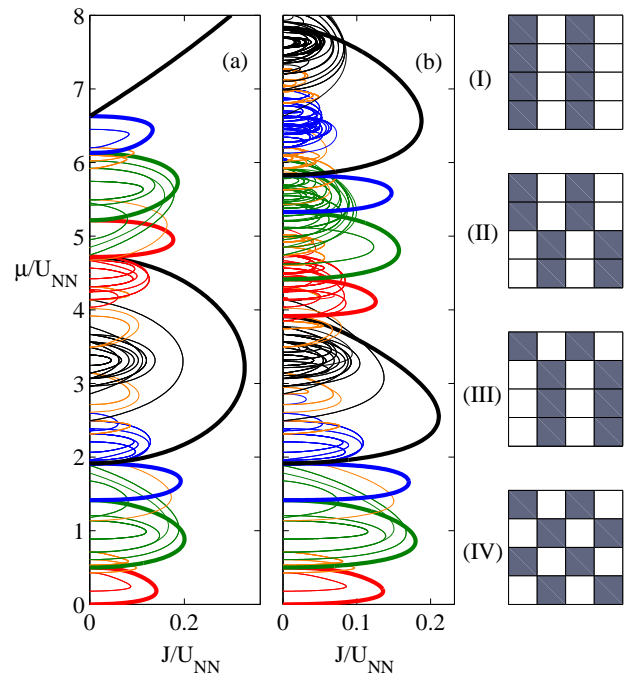}
\caption{(Color online) Phase diagram for weak and strong
dipole-dipole interaction: $U/U_{NN}=20$ (a) and $U/U_{NN}=2$ (b).
The thick lines are the ground state lobes, found (for increasing
chemicals potential) for filling factors equal to all multiples of
1/8. The thin lines of the same color are the metastable states at the
same filling factor. The other lines are for filling factors equal to
odd multiples of $1/8$ \cite{footnote2}; some of the metastable
configurations at filling factor 1/2 (I to III) and corresponding
ground state (IV). Empty sites are light and site occupied with 1 atom
are dark.}
\label{fig1}
\end{figure}
\end{center}

\vspace*{-0.75cm}

We study the problem in the mean field regime, with an approach based
on the Gutzwiller ansatz. This corresponds to writing the wavefunction
as a product over the different lattice sites $(i)$ of single-site
wavefunctions

\begin{eqnarray}
|\Phi(t)\rangle= \prod_i \sum_n f_n^{(i)}(t) |i,n\rangle.
\end{eqnarray}
In particular the time-depedence of the Gutzwiller coefficients
$f_n^{(i)}$ allows to study the evolution of the state in real ($t$)
and imaginary ($\tau=-it$) time

\begin{eqnarray}
i \frac{d\,f_n^{(i)}}{dt} &=& -J \left[ \bar{\varphi}_i \sqrt{n_i}
  f_{n-1}^{(i)} + \bar{\varphi}_i^* \sqrt{n_i+1} f_{n+1}^{(i)} \right] +   \\
&+& \left[\frac{U}{2} n_i(n_i-1) + \sum_{\vec \ell} U_{\vec \ell}\;
  \bar{n}_{i,{\vec \ell}} \; n_i -\mu n_i \right] f_n^{(i)}, \nonumber
\label{TDGA}
\end{eqnarray}
with $\varphi_i= \langle \Phi | a_i | \Phi \rangle$, $\bar{\varphi}_i=
\sum_{\langle j \rangle_i} \varphi_j$, $n_i=\langle \Phi | a^\dag_i
a_i | \Phi \rangle$, and $\bar{n}_{i,{\vec \ell}} = \sum_{\langle
\langle j \rangle\rangle_{i,{\vec \ell}}} n_j$.

The imaginary time evolution, which due to dissipation is supposed
to converge to the ground state of the system, in the presence of
long-range interaction happens to converge often to different
configurations, depending on the exact initial conditions. This is
a clear sign of the existence of {\it metastable} states in the
system. In the real time evolution, their stability manifests in
typical small oscillations at frequency $\omega_0$ around a local
minimum of the energy.  All the insulating metastable
configurations present an insulating lobe in the $J-\mu$ phase
space, as explained below.  They have a finite lifetime due to the
tunneling to different metastable states, which can be very long
for small tunneling parameter $J$ and large systems.  Using a path
integral approach in imaginary time \cite{wen}, combined with a
dynamical variational method (cf. \cite{perez}), we have estimated
the tunneling time $T$ to diverge for $J \to 0$ and to scale like
$\omega_0 T \approx \exp[N_s\hbar] \exp[-N_s \hbar J/{\tilde J}]$
for $J/{\tilde J} \gtrsim 0.3$, being $N_s$ the number of sites
and ${\tilde J}$ of the order of the tip of the lobe
\cite{trefzger}.

\begin{center}
\begin{figure}[t]
\includegraphics[width=0.9\linewidth]{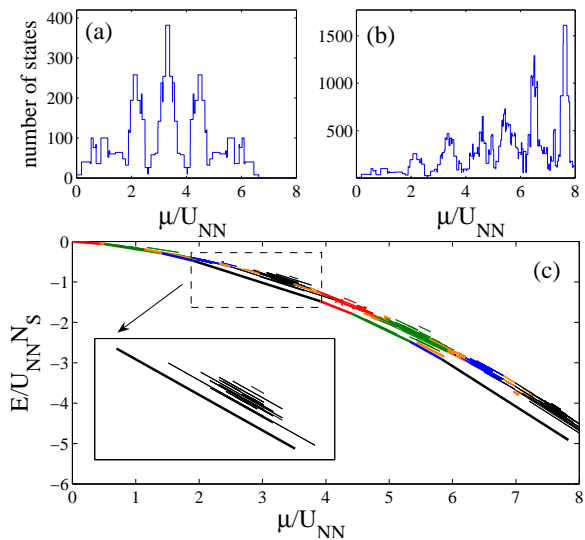}
\caption{(Color online) Number of metastable states as a function of
$\mu$ for weak and strong dipole-dipole interaction: (a)
$U/U_{NN}=20$ and (b) $U/U_{NN}=2$. (c) Energy of the ground (thick
line) and metastable states (thin lines) as function of $\mu$ for
strong dipole-dipole interaction ($U/U_{NN}=2$). The inset shows the
energy levels at filling factor $1/2$.}
\label{fig2}
\end{figure}
\end{center}

\vspace*{-0.75cm}

The most convenient method to determine the phase diagram of the
metastable states is to use a mean-field approach perturbative in
$\varphi_i$. Performing the mean field decoupling of the Hamiltonian,
the tunneling part at 1st order in the order parameter takes the form
$H_t=-J\sum_i (\bar{\varphi}_i^* a_i + \bar{\varphi}_i
a_i^{\dag})$. Using the definition $\varphi_i=\langle a_i e^{-\beta
H}\rangle$ in the limit $\beta \to \infty$, for a given classical
configuration described by the density distribution $n_i$, one gets

\begin{eqnarray}
\varphi_i=J \bar{\varphi_i}
\left[\frac{n_i+1}{Un_i-\mu+V^{1,i}_{dip}}-
\frac{n_i}{U(n_i-1)-\mu+V^{1,i}_{dip}}\right],
\label{pert}
\end{eqnarray}
where $V^{1,i}_{dip}$ is the dipole-dipole interaction of {\it one}
atom placed at site $i$ with the rest of the lattice.  A classical
configuration $n_i$ is defined {\it metastable }, if there exists a
region of the phase space $J - \mu$ (insulating lobe), where
Eq.(\ref{pert}) only allows the trivial solution $\varphi_i=0$,
$\forall i$.
The insulating lobes exactly coincide with the stability regions found
with the imaginary time approach. Studying the properties of
Eq.(\ref{pert}), one can determine in a reliable way the insulating
lobes for a huge number of configurations, which
% would be unaccessible
it would be impossible to access only by looking at the convergence of
the imaginary time evolution.

We investigate all possible configurations in $4 \times 4$ lattices
with periodic boundary conditions, with all filling factors $N_a/N_s$
(number of atoms/number of sites), ranging from $1/N_s$ up to one,
including the possibility of having double occupancy of the lattice
sites. The quantities of interest that we extract from this analysis
are: (i) the boundary of the lobes for the insulating configurations;
(ii) for each value of the chemical potential, the number of
metastable insulating states present at very low tunneling; (iii) the
energy of the ground state; (iv) the energy of all the insulating
metastable states.

Those results are summarised in Figs. \ref{fig1} and \ref{fig2}.  We
observe that for weak dipole-dipole interaction (similar to the hard
core limit) the system presents an almost exact particle-hole duality
(Figs.\ref{fig1},\ref{fig2}(a)), while for small on-site interaction,
which allows double occupation of the lattice sites, many more
configurations arise at filling factors larger than $1/2$
(Figs.\ref{fig1},\ref{fig2}(b)).  As shown in Fig.\ref{fig2}(c), there
is usually a gap between the ground state and the lowest metastable
state, which might allow to reach the ground state by ramping up the
optical lattice under some adiabaticity condition. However, this
feature is strongly reduced in the case of larger lattice sizes that
we are going to discuss in the following.

\begin{center}
\begin{figure}[t]
\includegraphics[width=0.925\linewidth]{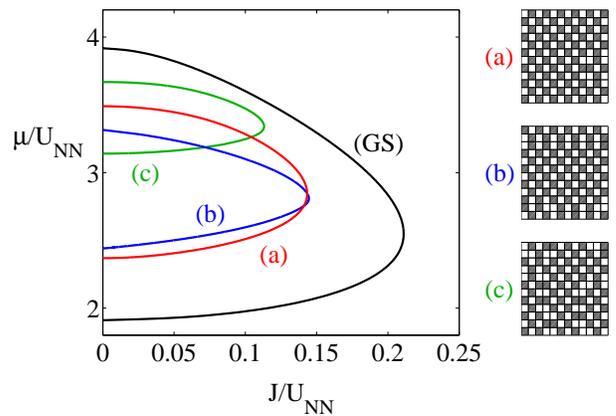}
%%%% created with fig_meta_NEW
\caption{(Color online) Phase diagram for the ground state at filling
factor $1/2$ and three metastable insulating configurations for
$U/U_{NN}=2$.}
\label{fig_meta_PD}
\end{figure}
\end{center}

\vspace*{-0.8cm}

The number of metastable configurations and the variety of their
patterns increase very rapidly with the lattice size. Since for
lattice sizes larger than $4 \times 4$ it is not possible to track
down systematically all existing configurations, we used a {\it
statistical} approach where we run many times the imaginary time
evolution for the same values of the parameters, each time changing
the initial conditions. Exactly for the same reason why the metastable
states exist, the convergence of such a procedure might be very very
slow and is not always accurate. Hence, the stability of each of the
obtained configurations is tested using the mean-field perturbative
approach described in Eq.(\ref{pert}), in order to confirm the
existence of an insulating lobe.  In general, we find that the
configurations which differ from very regular ones by small defects
are stable in a large region of the phase space, while the lobes
corresponding to configurations with many defects are very small. In
Fig.\ref{fig_meta_PD}, we show the insulating lobes for three
configurations which differ from the checkerboard only by small
defects.

\begin{center}
\begin{figure}[t]
\includegraphics[width=1\linewidth]{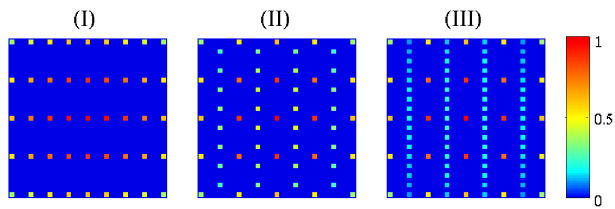}
%%%% created with fig_noise_rep
\caption{(Color online) Normalised spatial noise correlation patterns
for configurations (I) to (III) in Fig.\ref{fig1} \cite{footnote3}.}
\label{fig_fft2}
\end{figure}
\end{center}

\vspace*{-0.8cm}

Very important issues are the initialisation and detection of the
atomic states in the lattice. One can use superlattices in order to
prepare the atoms in configurations of preferential symmetry.
This idea is pursued by several experimental groups \cite{expe}.
We have checked that the presence of defects is strongly reduced when
a local potential energy following desired patterns is added to the
optical lattice. Note that the configurations obtained in such a way
will also result stable once the superlattice is removed, thanks to
dipole-dipole interaction.

The spatially modulated structures created in such a way can be
detected via the measurement of the noise correlations of the
expansion pictures \cite{altman,bloch,scarola}: the ordered structures
in the lattice give rise to different patterns in the spatial noise
correlation function, equal to the modulus square of the Fourier
transform of the density distribution in the lattice.  Such a
measurement is in principle able to recognize the defects in the
density distribution, which could be exactly reconstructed starting
from the patterns in the spatial noise correlation function. The
signal to noise ratio required for single defect recognition is beyond
the present experimental possibilities. However, averaging over a
finite number of different experimental runs producing the same
spatial distribution of atoms in the lattice, a good signal can be
obtained.
In Fig.\ref{fig_fft2}, we show the noise correlations for the
metastable configurations at filling factor $1/2$ shown in
Fig.\ref{fig1}, (I) to (III).

The capability of initialise and read-out the state of the lattice
makes those systems useful for applications as quantum memories.  The
controlled transfer of those systems from one configuration to another
will be object of future studies.

Alternatively to superlattices, structures obtained e.g. with atoms
chips or microlenses arrays, where each lattice site can be addressed
individually, could be used to prepare the desired configuration and
manipulate it.  Finally, it will be worth investigating the
possibility of creating an atom-light interface to initialize or
read-out the atomic state of the system by coupling it with the
polarisation degrees of freedom of light \cite{eckert}, for instance
exploiting the spinor character of Chromium dipolar atoms
\cite{sengupta2}.

We acknowledge financial support of ESF PESC "QUDEDIS", EU IP "SCALA",
Spanish MEC under contracts FIS 2005-04627 and Consolider-Ingenio 2010
"QOIT". C.M. acknowledges financial support by the EU under
Marie-Curie Fellowship. The authors thank L.~Santos and P.~Pedri for
interesting discussions.


\begin{thebibliography}{99}
\bibitem{footnote0} In the present work, we do not discuss the case of
a multi-layer structure of 2D planes (like the one obtained e.g. in a
3D optical lattice), where due to the long range of dipole-dipole
interaction the different planes are not independent even for
completely suppressed tunneling.
\bibitem{footnote} In standard experiments with $^{52}$Cr, $U/U_{NN}
\approx 400$ (for $\varepsilon_{dd}= \mu_0 (6\mu_B)^2 m/(12 \pi
\hbar^2 a) \approx 0.158$ and sperical localisation of the bottom of
the potential well at $s=20E_R$, where $E_R$ is the recoil energy at
$\lambda=500 \;{\rm nm}$). Using Feshbach resonances, $U/U_{NN}$ can
be tuned down to zero.
\bibitem{footnote2} In the imaginary time evolution, we find also
superfluid and supersolid phases, that we do not show in
Fig.\ref{fig1}.
\bibitem{footnote3} A localised gaussian density distribution at each
lattice site has been assumed to obtain these pictures.
\end{thebibliography}
\end{document}